\documentclass[a4paper]{article}

\usepackage{icrc2013}

\title{The Highest Recorded Proton Spectrum at Earth since the Beginning of the Space Age}

\shorttitle{Highest Recorded Proton Spectrum at Earth}

\authors{M.S. Potgieter$^{1}$, R. du T. Strauss$^{1}$, N. De Simone$^{2}$, 
M. Boezio$^{3}$
}

\afiliations{
$^1$Centre for Space Research, North-West University, Potchefstroom, South Africa\\ 
$^2$INFN, Sezione di Roma Tor Vergata, Rome, Italy\\ 
$^3$INFN, Sezione di Trieste, Trieste, Italy
}

\email{Marius.Potgieter@nwu.ac.za}

\abstract{ The recent solar minimum activity period and the consequent minimum modulation conditions for cosmic rays were unusual compared to previous solar minimum periods. 
The highest spectra of galactic protons (and other cosmic rays) were recorded by the PAMELA instrument at Earth in late 2009, in contrast to expectations. The spectrum, between 100 MeV and 50 GeV, for December 2009 is compared to proton spectra observed during previous solar minimum periods, back to 1965. Corresponding numerical modeling is presented which predicts that the next solar minimum spectra could even be higher if similar modulation conditions then would occur as in 2008-2009. The reason is that incorporating gradient and curvature drifts in modulation models causes proton spectra for A $>$ 0 solar magnetic cycles (e.g., around 1976, 1997) to always be higher than during A $<$ 0 cycles (e.g. around 1965, 1987, 2009) at energies below a few GeV, if the same modulation conditions would prevail.}


\begin{document}
\maketitle

\section{Introduction}

The past solar minimum activity period and the consequent minimum modulation conditions for galactic cosmic rays (CRs) were unusual. It was expected that the new activity cycle would begin early in 2008, assuming a 10.5 year periodicity. 
Instead, solar minimum modulation conditions had continued until the end of 2009, characterized by a much weaker heliospheric magnetic field (HMF) compared to previous cycles. 
The tilt angle of the wavy heliospheric current sheet (HCS), on the other hand, had not decreased as rapidly as the magnitude of the HMF at Earth during this period, but eventually also reached a minimum value at the end of 2009. It was 
reported by several groups that CRs with high rigidity reached record setting intensities during this time [1,2,3,4,5]. 
It has become evident that the last period of declining and eventually minimum solar activity, especially from mid-2006 to the end of 2009, 
and the subsequent increase in the CR intensity (differential flux) were different than the four previous cycles.

The highest modulated spectra between 100 MeV and 50 GeV of galactic protons (and other CRs) were recorded by the PAMELA instrument at Earth in late 2009 [6],
in contrast to expectations based on comprehensive modulation modelling as will also be shown here. 
We compare this spectrum to those observed between 30 MeV and 20 GeV during previous solar minimum periods, 
back to the solar minimum of 1965, and discuss why the high spectra of 2009 were unexpected. 

We also compare computed spectra to the PAMELA proton spectra observed from mid-2006 to December 2009 [6]. 
This model predicts that the next solar minimum CR spectra at Earth could even be higher if similar modulation conditions would then occur as in 2008-2009. 
The reason for this is discussed.

\begin{figure*}[!ht]
 \centering
 \includegraphics[width=0.65\textwidth]{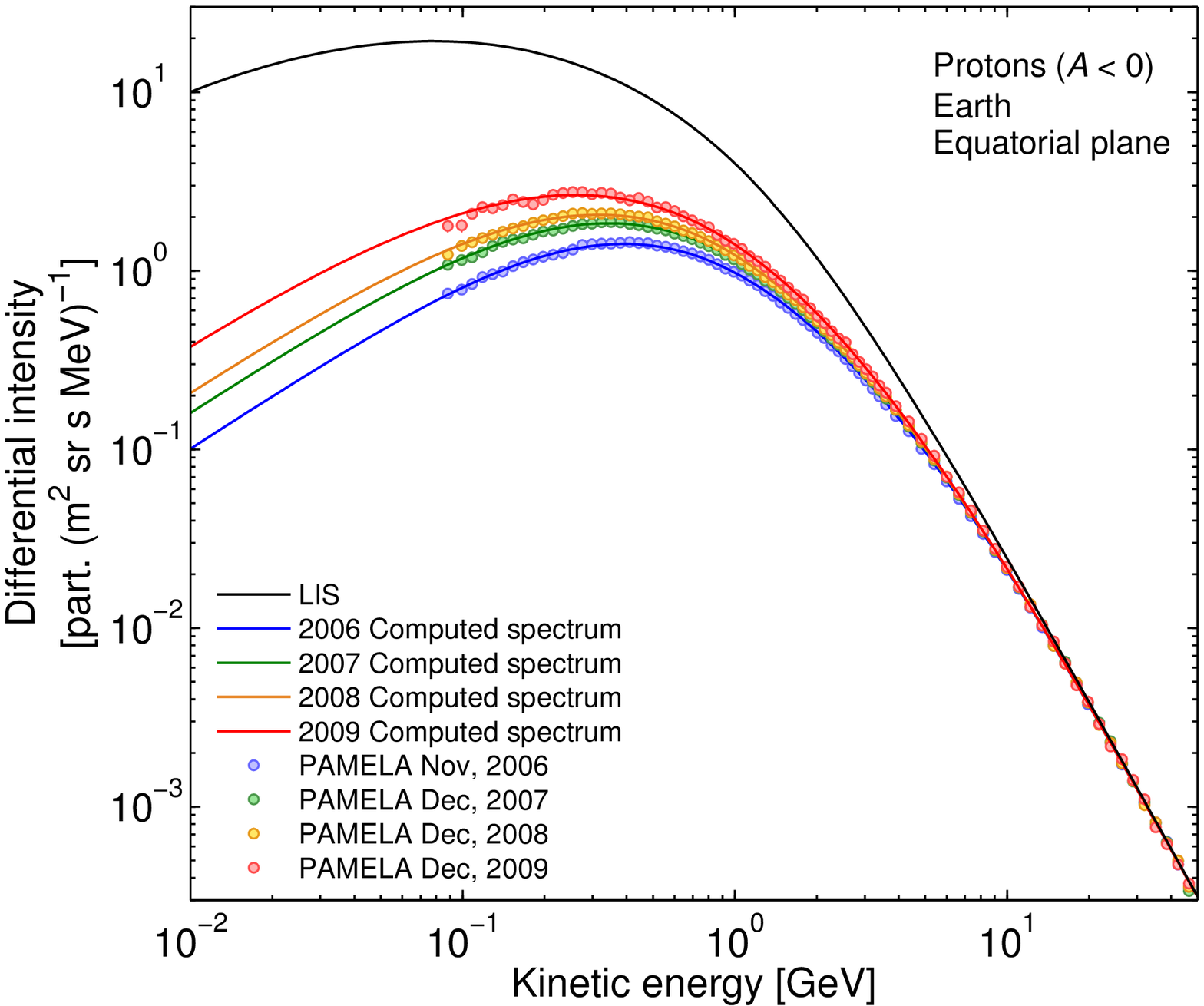}
 \caption{PAMELA proton spectra from 2006 (blue symbols) to 2009 (red symbols), overlaid by the corresponding computed spectra (solid lines). 
During this time the tilt angle of the heliospheric current sheet (HCS) changed from $\alpha = {15.7}^\circ$ to $\alpha = {10.0}^\circ$, 
with an accompanying change in the averaged HMF magnitude at Earth from $\it {B} \approx$ 5.05 nT to $\approx$ 3.94 nT . The LIS is specified at 120 AU where the modulation boundary is assumed to be located.
Figure adapted from [6,7,19]. }
\label{fig:Figure1}

\bigskip

\includegraphics[width=0.95\textwidth]{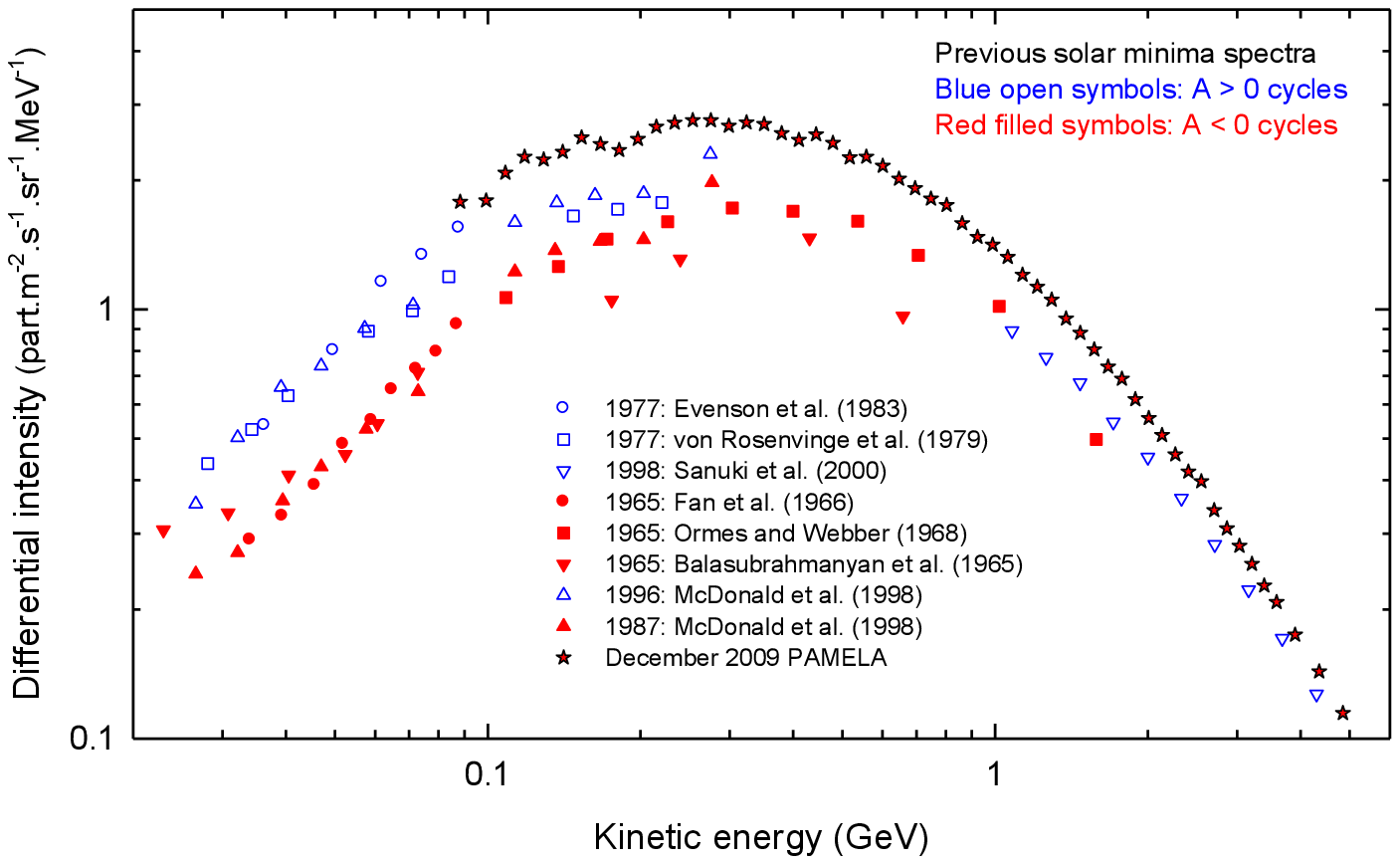}
 \caption{A comparison of CR proton spectra observed during solar minimum modulation periods of 1965, 1976-77, 1987, 1996-97 and 2009. Note that the spectra for all previous A $>$ 0 cycles 
(in blue) are consistently higher than for the A $<$ 0 cycles (in red) at energies below a few GeV, except for the 2009 spectrum that went up well above 
all the A $<$ 0 spectra and surprisingly also above the A $>$ 0 spectra. It is clearly the highest spectrum ever recorded since the beginning of the space age. Data are from [6,7,12,13,14,15,16,17]. }	
\label{fig:Figure2}
\end{figure*}

\section{Proton spectra for 2006 to 2009 compared to modelling results}

Potgieter et al. [7] utilized a comprehensive numerical modulation model to establish what mechanisms were exactly responsible for the 
modulation of protons from 2006 to solar minimum modulation in 2009, and why the observed proton spectrum for December 2009 was the highest modulated spectrum every recorded at Earth since the beginning of the space age. 
They used the extraordinary precise measurements of protons for this period by the PAMELA space experiment [6] to perform their study.
What they found is discussed further below.

A full three-dimensional (3D) modulation model, based on the numerical solution of the well-known heliospheric transport equation [8], was used to compute the differential intensity of 10 MeV to 50 GeV protons at Earth.  
They [7] used the customary HCS tilt angle and the observed HMF values at Earth as proxies for solar activity. 
The detailed comparison of this model to the PAMELA data is shown in figure \ref{fig:Figure1} based on these proxies for solar activity. 
During this time the averaged tilt angle of the HCS changed from $\alpha = {15.7}^\circ$ to $\alpha = {10.0}^\circ$, with an accompanying change in the averaged HMF magnitude at Earth from $\it {B} \approx$ 5.05 nT to $\approx$ 3.94 nT [7]. 
The latter had a significant effect on the global modulation of CRs. In their model the proton LIS is specified at 120 AU where the modulation boundary is assumingly located.

The decrease in $\it {B}$ was extraordinary large so that apart from drifts caused by a decreasing HCS waviness (decreasing tilt angles) towards solar minimum, global curvature and gradient drifts 
also became relatively larger, while the rigidity dependence of the diffusion process had become less to produce progressively softer spectra from 2006 to 2009. 
The softening of the modulated spectra at low energies could not be explained by simply allowing less overall diffusion towards solar minimum but required 
a significant weakening in the rigidity dependence of the major diffusion coefficients at rigidities below a few GV [7]. 
A similar observation was reported by Bazilevskaya et al. [5]. 

The modulation minimum period of 2009 can thus be described as relatively more ‘diffusion dominated’ than previous solar minima. 
However, drifts still had played a significant role but the observable modulation effects were not as well correlated with the waviness of the heliospheric current sheet as before. 
Protons still experienced global gradient and curvature drifts as the heliospheric magnetic field had decreased significantly until the end of 2009, 
in contrast to the moderate decreases observed during previous minimum periods. This caused a very fascinating interplay between all four major mechanisms during the period 2008-2009.

The 2009 solar minimum modulation period was unusual and clearly different than previous A $<$ 0 polarity minima.
For a review on drift effects, see [9].

\section{Comparison of five solar minimum spectra}

Next, the spectrum as shown in figure \ref{fig:Figure1} for December 2009 is compared to proton spectra observed during the previous four solar minimum epochs, 
that is, around 1965 (A $<$ 0), around 1976 (A $>$ 0) around 1986-7 (A $<$ 0) and around 1997-8 (A $>$ 0).

The present cycle (from one polarity change to the next) is called an A $<$ 0 polarity cycle when according to drift models, protons primarily drift to Earth via the equatorial 
regions of the heliosphere and by encountering the wavy HCS in the process, produce sharp intensity-time CR profiles. 
During A $>$ 0 cycles, protons drift to Earth primarily through the polar regions of the heliosphere and by subsequently missing the wavy HCS mostly, exhibit flattish profiles [see also 7,9]. 

The comparison is shown in figure \ref{fig:Figure2}. It is evident that the spectra for all previous A $>$ 0 cycles are consistently higher than for the A $<$ 0 cycles at energies below a few GeV, in accord with drift models. 
The 2009 spectrum (A $<$ 0 cycle), however, went up well above all the previous A $<$ 0 spectra and unexpectedly also above all the A $>$ 0 spectra. See also [20].
Evidently, the modulated spectrum for December 2009 is the highest proton spectrum ever recorded at Earth.

\section{Proton spectrum predicted for the next solar minimum}

We additionally applied a Stochastic Differential Equation (SDE) based modulation model by Strauss et al. [10,11] to compute and predict the proton spectrum for the next solar minimum period. 
This model is equivalent to the one used by [7] (except that the exact detail of the rigdity dependence of the diffusion coefficients as used in [7] for figure 1 was not implemented; 
the SDE model is numerically different but very stable with several advantages as illustrated by [10,11]).

\begin{figure}[!t]
\centering
			\includegraphics[width=0.5\textwidth]{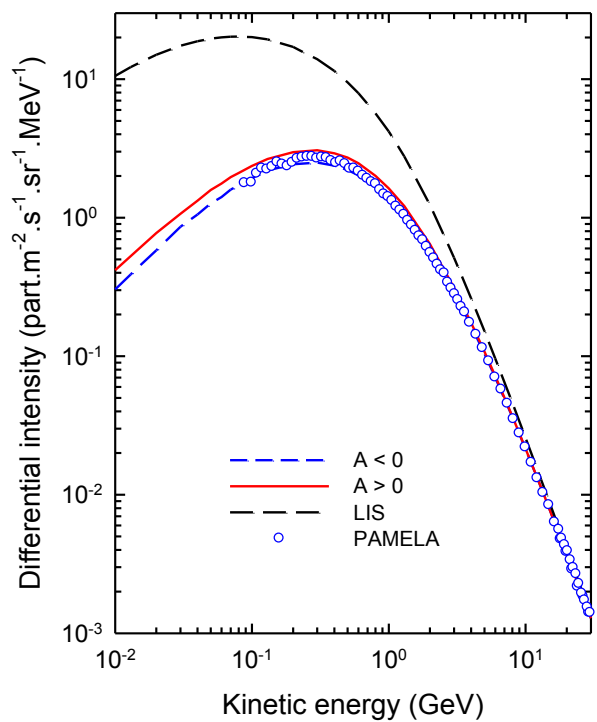}
			\caption{Computed modulated proton spectra at Earth for the two drift cycles (A $<$ 0 blue, lower dashed line; A $>$ 0 red, solid line) with respect to the local interstellar spectrum
(LIS; upper dashed line) specified at the modulation boundary, 
in comparison with the PAMELA observations during the last month of 2009 [6]. The last solar minimum was set in an A $<$ 0 cycle. This prediction for the next A $>$ 0 solar minimum
is based on the assumption that the same modulation conditions as in 2009 would prevail then. If this would happen the spectrum for the next solar minimum modulation period could become 
the highest every recorded proton spectrum.}
\label{fig:Figure3}
\end{figure}

The computed proton spectra are shown in figure \ref{fig:Figure3}, using the assumption that solar modulation conditions 
(meaning all proxies for solar activity as used in the modelling of solar modulation; see [7])
for the next solar minimum activity epoch will be identical to what had been observed between 2006 and 2009. 
First, it is evident that the A $<$ 0 spectrum from the SDE model is compatible with the observed 2009 spectrum, establishing the applicability of the SDE approach. 

Evidently, the modulated spectrum predicted for the next A $>$ 0 cycle is higher than the 2009 spectrum.
If the same modulation conditions as in 2009 would prevail during the next solar minimum, this may then become the highest every recorded proton spectrum and will set as such a new record.
If modulation conditions would enhance drifts during the next solar minimum, this predicted spectrum may even be higher.

\section{Conclusions}

Since the beginning of the space age, the highest modulated CR proton spectrum was observed by PAMELA in December 2009. 

This was unexpected because during previous A $<$ 0 polarity cycles, proton spectra were always lower than for A $>$ 0 cycles 
at kinetic energies less than a few-GeV, in full accord with drift models [e.g. 9,18].

The 2009 solar minimum modulation period was unusual and clearly different from previous A $<$ 0 polarity minima.

A comprehensive solar modulation model, including all four major processes and a full 3D tensor, is used to predict the proton spectrum for the next solar minimum 
and it was found that it could be higher than in 2009 and may therefore set a new record. The reason is that during A $>$ cycles 
it is easier for low energy particles to enter the inner heliopshere mainly through the polar regions of the heliosphere than in an A $<$ 0 cycles when they 
enter mainly through the equatorial regions. 

\vspace*{0.5cm}
\footnotesize{{\bf Acknowledgment:}{~Partial financial support of the South African National Research Foundation (NRF), 
and the SA High Performance Computing Centre (CHPC) is gratefully acknowledged.}}


\begin{thebibliography}{refs}

\bibitem{1} R.A. Mewaldt, et al., Astrophys. J. Lett. 273 (2010) L1-L6.
\bibitem{2} H.S. Ahluwalia and R.C. Ygbuhay, Adv. Space Res. 48 (2011) 61-64.
\bibitem{3} B. Heber, A. Kopp, J. Gieseler, R. Müller-Mellin, H. Fichtner, K. Scherer, M.S. Potgieter and S.E.S. Ferreira, Astrophysical J. 699  (2009) 1956-1963. 
\bibitem{4} F.B. McDonald, W.R. Webber and D.V. Reames, Geophys. Res. Lett. 37 (2010) L18101:1-5.
\bibitem{5} G.A. Bazilevskaya, M.B. Krainev, V.S. Makhmutov, Yu.I. Stozhkov, A.K. Svirzhevskaya and N.S. Svirzhevsky, Adv. Space Res. 49 (2012) 784–790.
\bibitem{6} O. Adriani, and PAMELA collaboration, Astrophys. J. 765 (2013) 91:1-8.
\bibitem{7} M.S. Potgieter, E.E. Vos, M. Boezio, N. De Simone, V. Di Felice and V. Formato, Solar Physics, in press (2013) arXiv:1302.1284.
\bibitem{8} E.N. Parker, Planet. Space Sci. 13 (1965) 9-49.
\bibitem{9} M.S. Potgieter, Adv. Space Res. (2013), in press. doi:10.1016/j.asr.2013.04.015.
\bibitem{10} R. du T. Strauss, M.S. Potgieter, I. Büsching and A. Kopp, Astrophys. J. 735 (2011) 83:1-13.
\bibitem{11} R. du T. Strauss, M.S. Potgieter, S.E.S. Ferreira, H. Fichtner and K. Scherer, Astrophys. J. Lett. 765 (2013) L18:1-6.
\bibitem{12} C.Y. Fan, G. Gloeckler and J.A. Simpson, Physical Rev. Lett. 17 (1966) 329-333.
\bibitem{13} J.F. Ormes and W.R. Webber, J. Geophys. Res. 73 (1968) 4231-4245.
\bibitem{14} T.T. von Rosenvinge, F.B. McDonald, J.H. Trainor and W.R. Webber, Proc. ICRC 12 (1979) 171-174.
\bibitem{15} T. Sanuki et al., Astrophys. J. 545 (2000) 1135-1142.
\bibitem{16}  P. Evenson, M. Garcia-Munoz, P. Meyer, K.R. Pyle and J.A. Simpson, Astrophys. J. Lett. 275 (1983) L15-18.
\bibitem{17}  F.B. McDonald, N. Lal and R.E. McGuire, J. Geophys. Res. 103 (1998) 373.
\bibitem{18}  D.M. Ngobeni and M.S. Potgieter, Adv. Space Res. 41 (2008) 373-380.
\bibitem{19}  E.E.Vos, Master’s thesis, North-West University, South Africa (2011).
\bibitem{20}  N. De Simone, Ph.D. thesis, University of Rome 'Tor Vergata', Italy (2011).

\end{thebibliography}
\end{document}